\documentclass[12pt]{iopart}
\usepackage{iopams}

\usepackage{graphicx} 

\def\be{\begin{equation}} 
\def\ee{\end{equation}} 
\def\ba{\begin{eqnarray}} 
\def\ea{\end{eqnarray}}

\def\ov{\overline} 
\def\I{{\rm Im}} 
 
\def\Z{\mathbb{Z}}

\def\nl{\nonumber \\}

\def\wh{\widehat} 
\def\Tr{{\rm Tr}}

\def\a{\alpha} 
\def\b{\beta}

\def\D{\Delta} 
\def\d{\delta}

\def\z{\zeta} 
 
\def\th{\theta} 
 
\def\l{\lambda} 
 
\def\m{\mu}

\def\p{\pi}

\def\s{\sigma} 
 
\def\t{\tau} 
\def\f{\phi}

\def\c{\chi}

\def\winfty{W_{1+\infty}}

\begin{document} 
 
 
\title[Coulomb Blockade in Hierarchical Quantum Hall Droplets]
{Coulomb Blockade in Hierarchical Quantum Hall Droplets} 
 
\author{Andrea Cappelli${}^1$, Lachezar S Georgiev${}^2$ and 
Guillermo R Zemba${}^3$} 
\address{${}^1$ INFN, Via G. Sansone 1, 50019 Sesto Fiorentino - 
Firenze, Italy}
\address{${}^2$ INRNE, Bulgarian Acad. of Sc., 
Tsarigradsko Chaussee 72,  1784 Sofia, Bulgaria}
\address{${}^3$ Facultad de Ciencias Fisicomatem\'aticas e Ingenier\'{\i}a,
UCA, Av. A. Moreau de Justo 1500,
(1107) Buenos Aires, Argentina and Departamento de F\'{\i}sica, CNEA,
Av.Libertador 8250, (1429) Buenos Aires, Argentina} 
\eads{\mailto{andrea.cappelli@fi.infn.it}, \mailto{lgeorg@inrne.bas.bg}, 
\mailto{zemba@tandar.cnea.gov.ar}}
 
\begin{abstract} 
The degeneracy of energy levels in a quantum dot of Hall 
fluid, leading to conductance peaks, can be readily derived from
the partition functions of conformal field theory. 
Their complete expressions can be found for Hall states with both  
Abelian and non-Abelian statistics, upon adapting known results for
the annulus geometry.  
We analyze the Abelian states with hierarchical filling fractions, 
$\nu=m/(mp\pm 1)$, and find a non trivial pattern of conductance peaks.
In particular, each one of them occurs with
a characteristic multiplicity, that is due to the extended symmetry of  
the $m$-folded edge. 
Experimental tests of the multiplicity can shed more light on the 
dynamics of this composite edge. 
\end{abstract} 

 
\section{Introduction} 
 
Among the recently proposed experimental tests of the quantum Hall effect
\cite{stern-rev}, the study of Coulomb blockade conductance peaks has been  
proposed in \cite{halp}\cite{stern}\cite{schou}. 
One considers an isolated droplet of Hall fluid that is formed in a
bar-shaped sample between two constrictions, in the limit of
strong quasiparticle backscattering.  
In this regime, electrons can tunnel into the droplet when energy states 
are led to be degenerate, either by changing the area of the dot
by means of a side modulation gate or by tuning the magnetic field.
 
Given that the low energy levels of Hall droplets,
the edge excitations \cite{wen}, are described by  
conformal field theory (CFT)  \cite{cft},
the level deformation and degeneracy can be obtained 
from the analysis of their known Hilbert spaces. 
The proper sector in these spaces is identified by the selection rules 
(the so-called fusion rules \cite{cft}) for the addition of  
electrons to the ground state. The presence of static 
quasi-particles in the interior of the droplet selects different sectors 
as dictated by the corresponding fusion rules. 
 
The following results were found in Refs.\cite{stern}\cite{schou}: 
 
i) In the simplest Laughlin Hall states \cite{das}, described by the 
chiral Luttinger liquid (chiral compactified free boson CFT) \cite{wen}, 
the peaks in the conductance as a function of the area $S$
are equally spaced by $\D S =e/n_o$, where $n_o$ is the electron density. 
The spacing is the same for any number of quasi-particles in 
the bulk of the droplet. 
 
ii) In the Read-Rezayi states \cite{rr}, with $\nu=2+k/(k+2)$, 
described by the chiral boson coupled to non-trivial neutral excitations 
of the $\Z_k$-Parafermion CFT \cite{cft}, the spacings between
peaks are modulated: a group of $k$ 
equidistant peaks at distance $\D S_1$ is separated 
from the next group by the spacing $\D S_2$, with $\D S_2>\D S_1$. 
The number of peaks in a group is lower in the presence of 
quasiparticles in the bulk and when relaxation (fusion) 
between bulk and boundary neutral excitations is allowed. 
 
The modulation follows from the energies
of the neutral excitations, that are coupled to the charged 
ones by a $\Z_k$ selection rule, and from the non-Abelian fusion rules.
These results led the authors of \cite{stern}\cite{schou} 
to conclude that the pattern of Coulomb blockade peaks 
provide signatures for the non-Abelian statistics of the excitations.
Analogous results were found in the case of the Ardonne-Schoutens 
non-Abelian spin singlet states \cite{ard}. 
 
In this paper, we discuss the case of the hierarchical Jain Hall states, 
with $ \nu=m/(mp\pm 1)$,  $m=2,3,\dots$, $p=2,4,\dots$, 
\cite{jain} that are described 
by the $m$-component Luttinger liquid and thus possess excitations 
with Abelian statistics \cite{hiera}. 
We show that the same type of modulation in 
groups of $m$ peaks occurs: however, it is independent 
of the presence of quasiparticles in the bulk  
and of bulk-edge relaxation phenomena (the latter are actually not possible).  
Furthermore, the peaks occur with a non-trivial pattern of 
multiplicities, owing to the extended symmetry of the multicomponent edge.

Our analysis shows that: 

i) Although the modulation of conductance peaks is not by itself 
a characteristic feature of non-Abelian statistics of excitations, 
the complete pattern of peaks and its dependence on the bulk
quasiparticles, provides an important experimental test
of the CFT description, in particular of the fusion rules and other 
qualitative properties of the Hilbert space.
 
ii) The characteristic degeneracy of each conductance peak is an interesting 
signature of the composite edge structure: from its lifting, one can check
additional interactions in the Hamiltonian,
such as irrelevant terms, inter-edge couplings, edge reconstruction effects 
etc. (see e.g. Ref. \cite{agam}).

\section{The partition function on the disk geometry}

We start by describing the 
edge excitations by means of the CFT partition function on the 
disk geometry (droplets with different shapes are equivalent, 
owing to the symmetry of the QHE under area-preserving deformations  
\cite {ctz}). 
The use of the (grand-canonical) partition function greatly  
simplifies the analysis: it not only accounts for all states in the theory 
but also for the selection rules which are built in. 
The partition functions for the prominent Hall states 
are already known in the annulus geometry \cite{cz}\cite{cgt}, from which
we shall deduce those for the case of the disk. 
 
We illustrate the method and set the notation by first repeating  
the analysis of the simpler Laughlin states, $\nu=1/q=1,1/3,1/5,\dots$. 
The CFT partition function describes excitations 
living on the two edges of the annulus and
having opposite chiralities; the
bulk is static and nothing depends on the radial coordinate.
The angular and time coordinates are both periodic, the latter 
being the inverse temperature $\b$:
therefore, the spacetime geometry is that of a torus.
The partition function is \cite{cz}:
\be 
Z=\sum_{\l=1}^{q}\ \left\vert\th_\l(\t,\z)\right\vert^2\ , 
\label{z-ann} 
\ee 
in terms of the (extended) conformal characters,
\be
\th_\l(\t,\z)=\Tr_{{\cal H}^{(\l)}}\left[\exp\left( 
i2\pi(\t (L_0 -c/24) +\z Q)\right)\right] \ .
\ee
Their expressions can be exactly computed from the representation theory
of the Virasoro algebra of conformal transformations 
and its extensions, including the current algebra
(affine Lie algebra) $\wh{U(1)}$.
Each character resums the states on one edge of the annulus, within
the Hilbert-space  sector ${\cal H}^{(\l)}$ with charges $Q=\l/q+\Z$. 
The parameters in the exponent are given by  $\I \t =v\b/2\pi R$, in terms 
of the inverse temperature, Fermi velocity and edge length $2\pi R$, 
and by $2\p\z=\b(\m +i V_o)$, in terms of the chemical and electric potentials, 
respectively. 
Therefore, states are collected according to their charge and energy:
the latter is proportional to the conformal dimension, $E =(v/R)(L_0 -c/24)$,
given by the zero mode $L_0$ of the Virasoro algebra, 
while the central charge $c$ accounts for the Casimir energy \cite{cft}. 

In (\ref{z-ann}), the sectors relative to the chiral excitations on 
the outer edge  of the annulus (accounted by $\th_\l$) are coupled  
to those of the inner edge ($\ov{\th}_\l$) such that only integer charge 
excitations exist globally. 
In $\ov{\th}_\l$, the variable $\ov{\t}$ may be rescaled w.r.t.
the complex conjugate of $\t$ for allowing different parameters on
the two edges: $\ov{v}/\ov{R}\neq v/R$.
Furthermore, we shall extend $\t$ to general complex values,
$\I \t >0$, such that the
evolution in $Z$ takes place both in time and space.

The annulus partition functions, such as (\ref{z-ann}), are given
by bilinear combinations of characters that obey 
the conditions of invariance under modular transformations. 
These are coordinate transformations that respect
the double periodicity of the torus \cite{cft}.
The most interesting transformation exchanges the two periods,
 $S:\ Z(\t,\z)=Z(-1/\t,-\z/\t)$. 
This is achieved in (\ref{z-ann}) by a linear transformation of the characters, 
$\th_\a(-1/\t)=\sum_\b\ S_{\a\b}\ \th_\b(\t)$,  
with $S_{\a\b}$ a unitary matrix. 
As discussed in  Ref.\cite{cz}, all modular conditions have 
physical interpretation in the Hall system.  
The $S$ invariance amounts to a completeness 
condition for the spectrum of the theory and it is well known that
the matrix $S_{\a\b}$ also determines the fusion rules of the CFT \cite{cft}. 
A general result is that the number of sectors $q$ 
in the partition function (\ref{z-ann}), i.e. the dimension of the $S$ 
matrix, is equal to the topological order of the Hall state 
\cite{wen}\cite{cz}. 
Two other modular conditions are, $T^2:\ Z(\t+2,\z)=Z(\t,\z)$ and
$U:\ Z(\t,\z+1)=Z(\t,\z)$, that respectively impose fermionic statistics and
integer charge for the excitations of the whole system 
on the two edges \cite{cz}.
 
The disk partition function can be uniquely obtained from (\ref{z-ann}) 
by letting the inner radius $\ov {R}\to 0$, such that only the 
ground state contribution remains, $\ov{\th}_\l\to \d_{\l,0}$. 
Therefore, we find:
\be 
Z_{\rm Disk}^{(0)}=\th_0\ . 
\label{z-disk} 
\ee 
If, however, there are quasiparticles in the bulk of charge 
$Q_{\rm Bulk}=-\a/q$, then the condition of total integer charge selects 
another sector, $Z_{\rm Disk}^{(\a)}=\th_\a$. 
In conclusion, the disk partition function is given 
by the chiral conformal character $\th_\a$, with index
selected by the bulk boundary conditions. 
 
\section{Coulomb blockade conductance peaks} 

We now proceed to compute the degeneracy of energy levels yielding
the conductance peaks. For the Laughlin states, $\nu=1/q$, 
the disk partition function is a sum of 
characters of the $\wh{U(1)}$ affine current algebra with Virasoro 
central charge $c=1$ and read, up to the constant
$\exp\left(-\nu\pi(\I\z)^2/\I\t\right)$ \cite{cz}: 
\be 
\th_\l=K_\l(\t,\z;q) \ ,
\label{z-laugh} 
\ee 
with,
\be
K_{\l}(\t,\z;q)=  
\frac{1}{\eta}\sum_{n=-\infty}^\infty \exp i2\pi 
\left[\t\frac{(nq+\l-\s)^2}{2q}+\z \frac{nq+\l}{q} \right] , 
\label{k-fun}
\ee
and $\eta(\t)$ is the Dedekind function \cite{cft}. 
In this expression, $n$ is the number of electrons added to the edge
and $\s =B\D S/\f_0$ is the deformation 
of the energy of edge excitations due to the variation 
$\D S$ of the area of the droplet. As explained in \cite{stern},
this causes an unbalance of charge at the edge and
an electrostatic energy cost.

On the other hand, for variations of the magnetic field $B$, the edge CFT is
sensitive to the flux change, $\d= \D \f/\f_o =S\D B/\f_o$, as follows:
the partition functions are deformed by $K_\l\ \to\ K_{\l+\d}$,
in (\ref{k-fun}) with $\s=0$ , i.e. the energies and charges of all
states are changed.
For one quantum of flux, $\d=1$, each $\l$-sector ($\th_\l$)
goes into the following one: this is the so-called spectral flow \cite{cz}.
For example, the ground state $\th_0$ becomes the one-anyon sector 
$\th_1$, meaning that the higher $B$ has induced a quasi-hole of 
charge $Q=-1/q$ in the bulk and the edge states have adjusted 
correspondingly \cite{geo}. 
Note that the spectral flow leaves the annulus partition function 
(\ref{z-ann}) invariant; it amounts to 
the last modular invariance, $V:\ Z(\t,\z+\t)=Z(\t,\z)$ \cite{cz}. 
The spectral flow encodes the Laughlin
argument for the exactness of the Hall current \cite{das}: 
after adding one quantum of flux, the system goes back to itself, but
a charge equal to the filling fraction has moved from one edge to the other.

Consider the Hall dot without quasiparticles in the bulk 
corresponding to the partition function (\ref{z-disk}):  
from the character $K_0$ (\ref{k-fun}), we can extract the energies 
and charges of the electron excitations
as the expressions multiplying $\t$ and $\z$, respectively. 
Upon deformation of the dot area, the ground state energy, 
$E\sim \s^2/2q$, $Q =0$, and that of the one-electron state,  
$E\sim (\s-q)^2/2q$, $Q=1$, become degenerate at $\s=q/2$. 
At this point one electron can tunnel into the dot causing a
conductance peak. 
Similar degeneracies occur between consecutive multi-electron states. 
Therefore, the Coulomb blockade peaks are equally separated by the distance
$\s =q$, corresponding to $ \D S=\f_o/(B\nu)=e/n_o $.
In the presence of quasiparticles in the bulk with 
charge $Q=-\l/q$, one should repeat the analysis using the
partition functions $\th_\l$ (\ref{z-laugh}): one obtains the same 
peak separations, because the energies in the different $\l$ sectors 
are related by shifts of $\s$. 

We have thus recovered the results of \cite{stern} 
from the study of the partition function: its decomposition into sectors 
makes it clear the allowed electron transitions. 
We conclude that the study of Coulomb blockade 
peaks provides insight into the qualitative
and quantitative structure of the CFT Hilbert space of edge excitations.
 
\section{Conductance peaks in hierarchical states} 

Let us now study the Coulomb blockade peaks in the hierarchical FQH states 
with filling fraction $\nu=m/(mp+1)$, $m=2,3,\dots$, $p=2,4,\dots$.  
The edge theory involves a $m$-component chiral
Luttinger liquid, i.e. a $c=m$ CFT, with a symmetric spectrum
of charges corresponding to the extended 
affine symmetry $\wh{U(1)}\times\wh{SU(m)}_1$ \cite{hiera}.
Our starting point is the  annulus partition function, 
which is again of the general form (\ref{z-ann}), but where now each sector 
contains a non-trivial combination of $\wh{U(1)}$ characters $K_\l$ 
(\ref{k-fun}) and the characters $\c_\a(\t)$ for the neutral $(m-1)$ 
components \cite{cz}. 
The latter have no dependence on $\z$ and describe the spectrum with 
 $\wh{SU(m)}_1$ symmetry (in this special case, the affine symmetry actually
leads to Abelian fractional statistics and fusion rules \cite{cft}).
There are $m$ neutral characters, obeying $\c_{\a+m}=\c_\a$; 
their leading low-energy behavior is,  
for $\I\t\to\infty$, 
\be 
\c_\a(\t)\sim\left({m\atop \a} \right) \exp i2\pi\t \left(\frac{v_n}{v}
\frac{\a(m-\a)}{2m} \right)\ +\ \cdots , 
\label{chi-sum} 
\ee 
with $\a=0,1,\dots,m-1$.
 In this expression, we allowed for a different velocity $v_n$ of 
neutral excitations.

The topological order of hierarchical Hall fluids is given by the 
denominator of the filling fraction $q=mp+1$ \cite{hiera}; therefore, there 
are $q$ terms in the annulus $Z$ and correspondingly $q$ possible 
disk partition functions $Z_{\rm Disk}^{(a)}=\th_a$, $a=0,1,\dots,q-1$. 
Their expressions are \cite{cz}: 
\be 
\th_a{(\t,\z)}=\sum_{\b=1}^m K_{ma+\b q} (\t,m\z;mq)\ \c_\b(\t)\ . 
\label{z-hier} 
\ee 
For example, in the $\nu=2/5$ case, there are two neutral characters 
that combine with ten charged ones to obtain the following 
five sectors: 
\ba 
\th_0 &=& K_0 (\t,2\z;10)\ \c_0 +K_5(\t,2\z;10)\ \c_1\ ,\nl 
\th_{\pm 1} &=& K_{\pm 2} (\t,2\z;10)\ \c_0 +K_{5\pm 2} (\t,2\z;10)\ \c_1\ ,\nl 
\th_{\pm 2} &=& K_{\pm 4} (\t,2\z;10)\ \c_0 +K_{5\pm 4} (\t,2\z;10)\ \c_1\ . 
\label{z-ex} 
\ea 
We now search for degeneracy of energy levels
differing by the addition of one electron, $\D Q=1$. 
Consider for definiteness the $\nu=2/5$ case 
without any bulk quasiparticle, i.e. $\th_0$ above. 
From the expressions (\ref{k-fun},\ref{z-hier},\ref{z-ex}), 
one finds that the first term 
$K_0$  resumes all even integer charged excitations, while 
$K_5$ the odd integer ones. Therefore, the first conductance peak is 
found when the lowest energy state in $K_0\ \c_0$, i.e. 
the ground state, with $E=(v_c/R) (2\s)^2/20$, $Q=0$, 
becomes degenerate with the lowest one in $K_5\ \c_1$, with 
$E=(v_c/R)(-5 +2 \s)^2/20 +v_n/4R $, $Q=1$. 
The next peak occurs when the latter becomes degenerate with  
the first excited state ($Q=2$) in $K_0\ \c_0$, and so on. 
Note that in these energies, we changed $\s\to 2\s$,
(cf. (\ref{k-fun})), in order to respect the flux-charge relation, 
$\D Q =\nu \D\f/\f_o$.
Owing to the contribution of the neutral energy in $\c_1$ 
(cf. (\ref{chi-sum})), the level matching is not midway and there is 
a bunching  of peaks in pairs, with separations
$\s =5/2 \mp v_n/2v$.
 
For general $m$ values, the result can be similarly obtained from  
(\ref{chi-sum},\ref{z-hier}). 
Our analysis is close to that of \cite{stern}, because the neutral energies in
(\ref{chi-sum}) are equal, up to a factor of two, to those of the  
$\Z_m$ parafermions in the Read-Rezayi Halls states.  
For $\nu=m/q$, the resulting separation $\s_k$
between the $k$-th and $(k+1)$-th peaks reads ($\s=B\D S/\f_o$):
\ba
\s_k \! &=& \!  \frac{q}{m}- \frac{v_n}{v}\frac{1}{m}\ ,
\qquad\qquad d_k=\left({m\atop k}\right), \  k=1,\dots,m-1,
\cr
\s_m \!  &=& \!  \frac{q}{m}+ \frac{v_n}{v}\frac{m-1}{m}\ ,\ \qquad d_m=1\ . 
\label{peak} 
\ea 
The pattern repeats with periodicity $m$; in (\ref{peak}),
we also report the degeneracy $d_k$ for the $k$-th peak arising from
the multiplicity factor in (\ref{chi-sum}) (more on this later). 
Note that the result (\ref{peak}) could also be obtained from
the analysis of the $m$-dimensional lattice of excitations \cite{hiera},
the multiplicity being given by the set of shortest vectors with 
integer charge $k$ \cite{cgvz}.

The formulae (\ref{peak}) also hold for 
the hierarchical states with $\nu=m/(mp-1)$, upon replacing $q=mp-1$: 
actually, the corresponding partition functions are of the form  
(\ref{z-hier}) with the replacement, $\c_\b\to\ov{\c}_\b$ \cite{cz},
that does not affect the earlier discussion of energetics.  

We thus found that the modulation of Coulomb blockade peaks is also
possible in Abelian Hall states;
the hierarchical states behave nevertheless differently from the  
non-Abelian Read-Rezayi states in the following aspects: 
 
i) Most importantly, the neutral states (\ref{chi-sum}) occur with a
characteristic multiplicity $d_k$ in (\ref{peak}), which is due 
to the $\wh{SU(m)}_1$ symmetry. 
This means that several states become degenerate at the same point and
$d_k$ electrons tunnel into the dot simultaneously. 
Of course the CFT description applies to very large dots: in practice,
this degeneracy is lifted by finite-size effect. Thus, the conductance
peaks will have a comb-like substructure that could be observable by achieving
higher experimental resolution.

ii) The pattern of peaks (\ref{peak}) is the same 
for any number of quasiparticles in the bulk, because the sectors, 
$\th_a$, $a\neq 0$, have linearly shifted energies w.r.t those of $\th_0$ 
 (cf.(\ref{z-hier})) \cite{cgvz}. 
Note, however, that bulk quasiparticles can have multiplicities,
due to their neutral parts: e.g., for $\nu=2/5$ 
there are two quasiparticles with $Q=1/5$ (cf. (\ref{z-ex})). 
When such multiplicity is $\left({m \atop k-1} \right)$,
the sequence of peaks (\ref{peak}) starts from $\s_k$ (instead of $\s_1$)
and goes on periodically.
 
iii) The relaxation processes between edge and bulk excitations are 
not possible. As explained in Ref.\cite{schou}, the added electron 
at the boundary may decay into another excitation with same charge 
but different neutral content, by fusing with a neutral bulk quasiparticle.
In the hierarchical states, any charged component, 
$K_\l$, appears only once in the spectrum (cf.(\ref{z-ex})), 
thus relaxations cannot take place. 
 
\section{Discussion}

The dynamics of the $m$-composite edge has been 
the subject of intense debate in the recent literature, starting from 
the experimental result \cite{chang}. Several 
deformations of, or additions to the Luttinger liquid  Hamiltonian, 
have been put forward \cite{agam};
as these break the $\wh{SU(m)}_1$ symmetry and possibly the conformal symmetry,
they should lift the peak degeneracy. 
 
Here we would like to recall the proposal of the minimal 
$\winfty$ models \cite{w-min}: in these theories, the degeneracy is  
completely eliminated and conformal symmetry is restored. 
As described in Ref.\cite{h-red}, a (non-local) projection in the 
edge Hamiltonian removes the $\wh{SU(m)}_1$ symmetry while 
leaving the main feature of incompressibility of the Hall fluid, 
the so-called area-preserving of $\winfty$ symmetry \cite{ctz}. 
The conductance peaks in the minimal $\winfty$
theories are easily obtained: since the projection preserves the 
structure of the Hilbert space of the CFT, it does not modify the 
earlier expressions for the partition functions, but only replaces the 
neutral characters by other expressions whose leading terms (\ref{chi-sum}) 
do not contain the multiplicity factor \cite{h-red}. 
We conclude that the observation of the conductance peaks without 
any multiplicity may provide a check for the $\winfty$ minimal models. 
Further applications of disk partition functions 
will be given in a forthcoming publication \cite{cgvz}.

\ack

 We thank K. Schoutens, A. Stern and G. Viola for useful discussions. 
A.C. and L.S.G. would like to thank the hospitality of the G. Galilei 
Institute for Theoretical Physics, Florence.
L.S.G. has been supported by the Alexander 
von Humboldt Foundation and by BG-NSF under contract DO 02-257.
L.S.G. and G.R.Z. acknowledge INFN and the Department of Physics, 
University of Florence for partial support. 
G.R.Z. is a fellow of CONICET.
This work was done with the help of the ESF programme INSTANS and of a 
PRIN grant (Italy).

\end{document}